\documentclass[runningheads]{llncs}
\usepackage[T1]{fontenc}
\usepackage{latexsym}
\usepackage{caption}
\usepackage{subcaption}
\usepackage{graphicx}

\usepackage{xcolor}

\newcommand\blfootnote[1]{%
 \begingroup
  \renewcommand\thefootnote{}\footnote{#1}%
  \addtocounter{footnote}{-1}%
 \endgroup
}

\begin{document}
\title{Dynamic Data-Driven Digital Twins for Blockchain Systems}
%
%
\author{Georgios Diamantopoulos\inst{1,2} \and
Nikos Tziritas \inst{3} \and
Rami Bahsoon \inst{1} \and
Georgios Theodoropoulos \inst{2}*
}
\authorrunning{G. Diamantopoulos et al.}
%
\institute{School of Computer Science, University of Birmingham, Birmingham, United Kingdom \and
Department of Computer Science and Engineering and Research Institute for Trustworthy Autonomous Systems, Southern University of Science and Technology (SUSTech), Shenzhen, China \and
Department of Informatics and Telecommunications, University of Thessaly, Greece
}
\maketitle              
\begin{abstract}
In recent years, we have seen an increase in the adoption of blockchain-based systems in non-financial applications, looking to benefit from what the technology has to offer. Although many fields have managed to include blockchain in their core functionalities, the adoption of blockchain, in general, is constrained by the so-called trilemma trade-off between decentralization, scalability, and security. In our previous work, we have shown that using a digital twin for dynamically managing blockchain systems during runtime can be effective in managing the trilemma trade-off. Our Digital Twin leverages DDDAS feedback loop, which is responsible for getting the data from the system to the digital twin, conducting optimisation, and updating the physical system. This paper examines how leveraging DDDAS feedback loop can support the optimisation component of the trilemma benefiting from  Reinforcement Learning agent and a simulation component to augment the quality of the learned model while reducing the computational overhead required for decision making.

\end{abstract}

\blfootnote{*Corresponding Author}

\vspace{-30pt}
\section{Introduction}

Blockchain's rise in popularity is undeniable; many non-financial applications have adopted the technology for its increased transparency, security and decentralisation \cite{mansfield2017beyond}. Supply chain, e-government, energy management, IoT  \cite{supChain,egov,energy,IoT} are among the many systems benefiting from blockchain. 


Two main types of blockchain exist, namely, Public and Private  \cite{pubVpriv}  with Consortium \cite{consortium} being a hybrid of the two. From the above, private blockchain systems lend themselves easier to a dynamic control system. In a private blockchain, participating nodes, communicate through a peer-to-peer(P2P) network and hold a personal (local) ledger storing the transactions that take place in the system. This network is private and only identified users can participate. The set of individual ledgers can be viewed as a distributed ledger, denoting the global state of the system. A consensus protocol is used to aid in validating and ordering the transactions and through it, a special set of nodes called block producers, vote on the order and validity of recent transactions, and arrange them in blocks. These blocks are then broadcasted to the system, for the rest of the nodes to update their local ledger accordingly. With the above working correctly, the nodes of the system vote on the new state of the ledger, and under the condition of a majority, each local ledger, and thus the global state of the system, is updated to match the agreed new system state. The above eliminates the need for a central authority to update the system state and assures complete transparency.

Despite the potential of Blockchain in many different domains, factors such as low scalability and high latency have limited  the technology's adoption, especially in time-critical applications, while in general, blockchain suffers from the so-called trilemma trade-off that is between decentralisation, scalability, and security \cite{zhou2020solutions}. 


The most notable factor affecting the performance of the blockchain, excluding external factors we cannot control such as the system architecture, network, and workload, is the consensus protocol, with system parameters such as block time, and block interval getting a close second. The trilemma trade-off in combination with blockchains time-varying workloads makes the creation of robust, general consensus protocols extremely challenging if not impossible, creating a need for other solutions \cite{Consensus}. Although no general consensus protocol exists, existing consensus protocols perform best under specific system conditions \cite{bf,pbft,zyz,quorum}. Additionally, blockchain systems support and are influenced by dynamic changes to the system parameters (including the consensus protocol) during runtime. Thus there is a need for dynamic management of blockchain systems.   




Digital Twins and DDDAS have been utilised in autonomic management of computational infrastructures~\cite{10.5555/2429759.2429956,ONOLAJA20101241,FANIYI20121167,6838768} and the last few years have witnessed several efforts to bring together Blockchain and Digital Twins. However, efforts have focused on utilising the former to support the latter; a comprehensive survey is provided in ~\cite{10.1145/3517189}. 
Similarly, in the context of DDDAS, Blockchain technology has been utilised to support different aspects of DDDAS operations and components~\cite{9004727,Ronghua,Ronghua1}.

In our previous work \cite{diamantopoulos2022digital}, we presented  a Digital Twin architecture for the dynamic management of blockchain systems focusing on the optimisation of the trilemma trade-off and we demonstrated its use to optimise a blockchain system for latency. 
The novel contribution of this work is enriching Digital Twins design for blockchain-based systems with DDDAS-inspired feedback loop. We explore how DDDAS feedback loop principles can support the design of  info-symbiotic link connecting the blockchain system with the simulation and analytic environment to dynamically manage the trilemma. 
As part of the loop, we contribute to a control mechanism that uses Reinforcement Learning agent(RL) and combined with our existing simulation framework. The combination  overcomes the limitations of just relying on RL while relaxing the computational overhead required when relying solely on simulation.


The rest of the paper is structured as follows: Section \ref{Digital Twins for  Blockchain Systems} discusses the  utilisation of Digital Twins for the management of Blockchain systems and provides an overview of a Digital Twin framework for this purpose. Section  \ref{The DDDAS Feedback Loop} delves into the DDDAS feedback loop at the core of the Digital Twin and examines its different components. As part of the loop, it proposes a novel optimisation approach based on the combination of an RL and what-if analysis. Section \ref{Experimental setup and Evaluation} presents a quantitative analysis of the proposed optimisation approach. Finally, section \ref{Conclusions} concludes this paper.

\section{Digital Twins for  Blockchain Systems}
\label{Digital Twins for  Blockchain Systems}



For this paper, we consider a generic permissioned blockchain system illustrated as 'Physical System' in Fig. \ref{DT} with  K nodes denoted as: 

\begin{equation}
    P = \{p_1, p_2, ..., p_K\}
\end{equation}

\noindent M of which are block producers denoted as:

\begin{equation}
    B = \{b_1, b_2, ..., b_M\},  \ B \subset P
\end{equation}

\noindent which take part in the Consensus Protocol (CP) and are responsible for producing the blocks \cite{diamantopoulos2022digital}. Additionally, each node $p\in P$ holds a local copy of the Blockchain(BC) while the block producers $b\in B$ also hold a transaction pool(TP) which stores broadcasted transactions. 


\vspace{-10pt}
\subsection{Consensus} 
In the above-described system, nodes produce transactions, which are broadcasted to the system, and stored by the block producers in their individual transaction pools. Each block producer takes turns producing and proposing blocks in a round-robin fashion. Specifically, when it's the turn of a block producer to produce a  new block, it first gathers the oldest transactions in the pool, verifies them and packs them into a block, signs the block with its private key and initiates the consensus protocol. The consensus protocol acts as a voting mechanism for nodes to vote on the new state of the system, the new block in this case, and as mentioned earlier, is the main factor affecting the performance of the blockchain. It is pertinent to note that producing blocks in a round robin fashion is simple to implement albeit inefficient due to "missed cycles" caused by invalid blocks or offline nodes \cite{proposal}. Other alternative implementations are possible, such as having every block producer produce a new block or leaving the selection up to the digital twin.

Although consensus protocols have been studied for many years, due to their use in traditional distributed systems for replication, blockchains larger scale, in combination with the unknown network characteristics of the nodes, make the vast majority of existing work incompatible. Recent works have focused on adapting some of these traditional protocols for the permissioned blockchain, achieving good performance but so far no one has managed to create a protocol achieving good performance under every possible system configuration \cite{specConsensus}. With several specialized consensus protocols available, a dynamic control system is a natural solution for taking advantage of many specialised solutions while avoiding their shortcomings. 

The idea of trying to take advantage of many consensus protocols is not new, similar concepts already exist in the literature, in the form of hybrid consensus algorithms \cite{PoWverif,forkFreePoA} which combine 2 protocols in one to get both benefits of both. Although fairly successful in achieving their goal of getting the benefits of two consensus protocols, hybrid algorithms also combine the shortcomings of the algorithms and are usually lacking in performance or energy consumption. 
In contrast, a dynamic control system allows for the exploitation of the benefits of the algorithms involved, with the cost of additional complexity in the form of the selection mechanism.

In our previous work \cite{diamantopoulos2022digital}, we focused on minimizing latency by dynamically changing between 2 consensus protocols Practical Byzantine Fault Tolerance (PBFT) \cite{pbft} and BigFoot 
\cite{bf}. PBFT acts as a robust protocol capable of efficiently achieving consensus when byzantine behaviour is detected in the system while BigFoot is a fast alternative when there are no signs of byzantine behaviour \cite{bf}.

\begin{figure}[htp]
\includegraphics[width=0.95\textwidth]{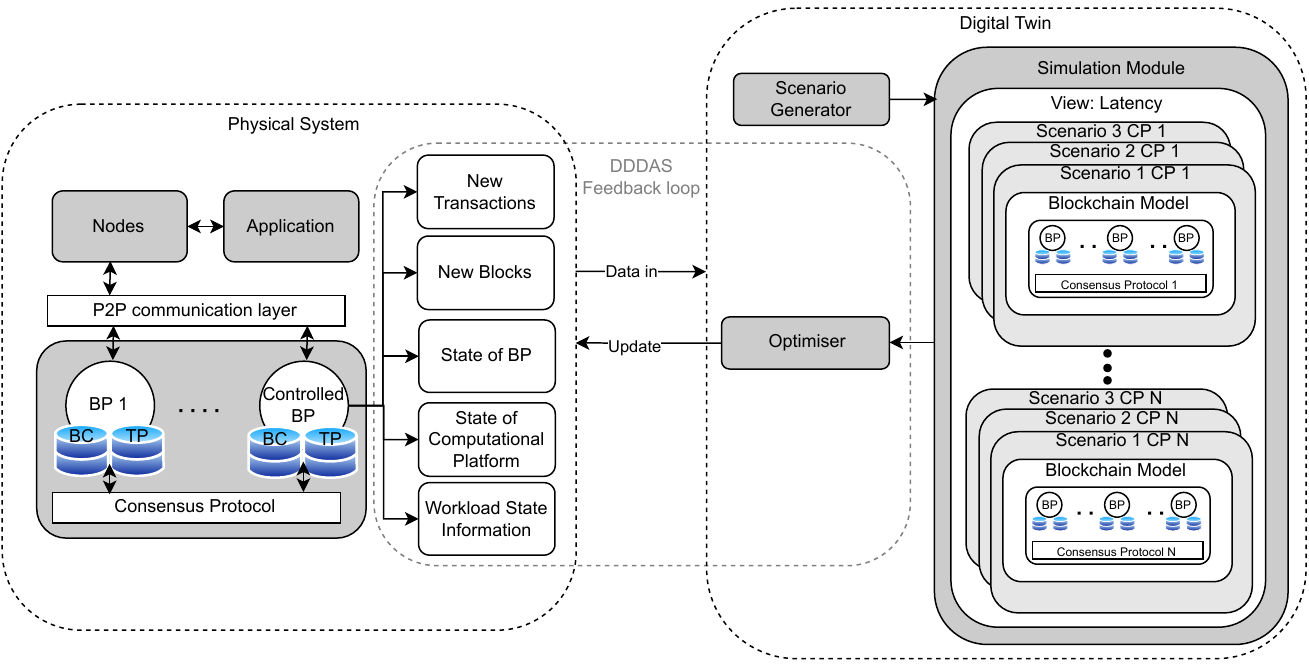}
\centering
\caption{Digital Twin Architecture and DDDAS feedback loop } 
\label{DT}
\end{figure}

To achieve the above, we employed a Digital Twin (DT) coupled with a Simulation Module to conduct what-if analysis, based on which, an optimiser would compute the optimal consensus protocol for the next time step. Using a DT can overcome the shortcomings of relying on an RL agent alone since the simulation element and what-if analysis allow for the exploration of alternative future scenarios \cite{whatIfAnalysis}. The complete architecture can be seen in Fig. \ref{DT}.

\section{The DDDAS Feedback Loop}
\label{The DDDAS Feedback Loop}

The system described in the previous section closely follows the DDDAS paradigm, with the Digital Twin containing the simulation of the physical system, the node feeding data to the Digital Twin acting as the sensors, and the optimiser updating the system closing the feedback loop. 

\subsubsection{Interacting with the blockchain system.}

Blockchains design, in combination with the communication protocol for the consensus, allows block producers to have access to or infer with high accuracy, a large amount of data about the state of the blockchain system. These block producers can act as the sensors of the physical system tasked with periodically sending state data to the digital twin. Specifically, every new transaction and block are timestamped and broadcasted to the system and thus are easily accessible. Using the list of historical transactions, we can develop a model of the workload used in the simulation. Although using queries to request the state of block producers requires a mechanism to overcome the Byzantine node assumption, blocks contain a large amount of data which could make the above obsolete. Each new block contains an extra data field in which the full timestamped history of the consensus process is stored and can be used to infer the state of the Block producers. Specifically, through blocks, we can learn the state of block producers (offline/online), based on whether the node participated in the consensus protocol or not, as well as develop a model of the block producers and their failure frequency. Additionally, using the relative response times we can infer a node's network state, and update it over time. With all of the above, a fairly accurate simulation of the blockchain system can be achieved.

\vspace{-5pt}
\subsubsection{Updating the model and controlling the physical system.} 
Relying on simulation to calculate the optimal system parameters is a computationally expensive approach \cite{diamantopoulos2022digital}. As the optimisation tasks get more complicated, with multiple views taken into account (figure \ref{DT}), smart contract simulation, harder to predict workloads, and especially once the decision-making process gets decentralised and replicated over many block producers, conducting what-if analysis becomes taxing on the hardware. Depending on the case i.e energy aware systems or systems relying on low-powered/battery-powered nodes might not be able to justify such an expensive process or worst case, the cost of optimisation can start to outweigh the gains. 

\subsection{Augmenting Reinforcement Learning with Simulation}

In this paper, we propose the use of a Reinforcement Learning (RL) agent  in combination with simulation and what-if analysis  to overcome the individual shortcomings of each respective technique. Reinforcement Learning trained on historical data cannot, on its own, provide a nonlinear extrapolation of future scenarios, essential in modelling complex systems such as blockchain \cite{rl}, while simulation can be computationally expensive. By using the simulation module to augment the training with what-if generated decisions the agent can learn a more complete model of the system improving the performance of the agent. Additionally, what-if analysis can be used when the agent encounters previously unseen scenarios, avoiding the risk of bad decisions. 
\begin{figure}
\includegraphics[width=.8\textwidth]{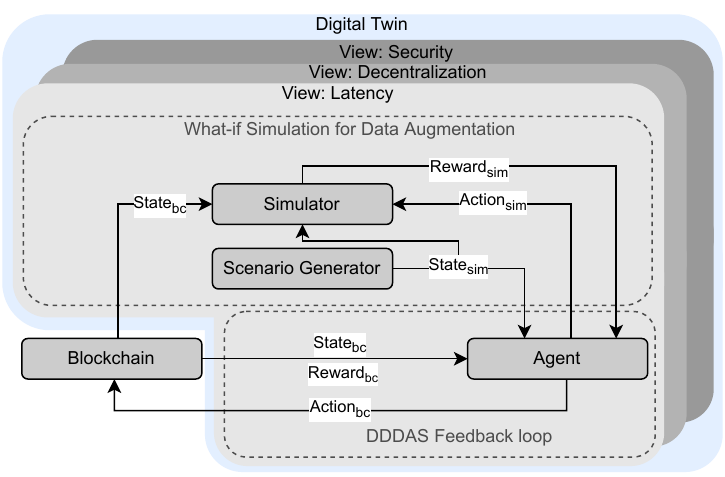}
\centering
\caption{General architecture for RL based control} 
\label{opt}
\end{figure}

For the optimisation of the trilemma trade-off, views are utilised with each view specialised in a different aspect of the trilemma \cite{diamantopoulos2022digital}. 
In this case, the DDDAS component may be viewed as consisting of multiple feedback loops  one for each aspect of optimisation. By splitting the DDDAS into multiple feedback loops, we can allow for finer control of both the data needed and the frequency of the updates. Additionally, moving away from a monolithic architecture allows for a more flexible, and scalable architecture. Specifically, each view consists of two components: the DDDAS feedback loop and the training data augmentation loop. The DDDAS feedback loop contains the RL agent which is used to update the system. The what-if simulation component includes the simulation module (or simulator) and the Scenario Generator. The data gathered from the physical system are used to update the simulation model while the scenario generator generates what-if scenarios, which are evaluated and used in the training of the agent. In Fig. \ref{opt} a high-level architecture of the proposed system can be seen.


\section{Experimental setup and Evaluation}
\label{Experimental setup and Evaluation}

\subsubsection{Experimental setup}

To illustrate the utilisation of RL-based optimisation and analyse the impact  of using simulation to enhance the RL agent, a prototype implementation of the system presented in figure \ref{opt} has been developed focusing on latency optimisation.  More specifically we consider  the average transaction latency defined as $\frac{\sum_{i}^{T_B}{Time_B - Time_{T_i}}}{T_B}$,
with $T_B$ denoting the number of transactions in the block $B$, $T_i$ the $i_{th}$ transaction in $B$ and $Time_B$, $Time_{T_i}$ the time $B$ and $T_i$ were added to the system, respectively.
\begin{figure}[h]
\includegraphics[width=.7\textwidth]{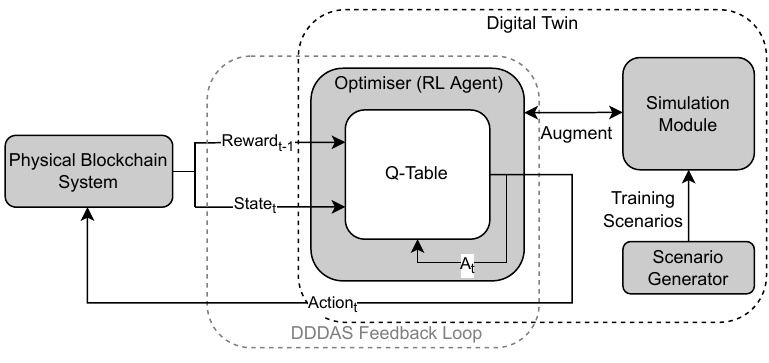}
\centering
\caption{An example instantiation of RL-based control: Latency View} 
\label{opt}
\end{figure}


For the experiments, a general permissioned blockchain system like the one described as "Physical System" in Fig. \ref{DT},  was used with 5 nodes taking part in the consensus protocol. Two consensus algorithms were implemented specifically, PBFT and BigFoot which complement each other as shown in \cite{diamantopoulos2022digital}. The scenario generator created instances of the above blockchain system by randomly generating values for the following parameters: (a) Transactions Per Second ($TPS$) which denotes the number of transactions the nodes generate per second; (b) $T$ which denotes the size of the transactions; (c) Node State which signifies when and for how long nodes go offline; and (d) Network State which denotes how the network state fluctuates over time. Additionally, following our previous approach, we assume that the system does not change state randomly, but does so in time intervals of length $TI$. Finally, the digital twin updates the system in regular time steps of size $TS$. 

A Q-Learning agent has been used. The state of the system $S$ is defined as $S = (F, N_{L}, N_{H})$ with $F$ being a binary metric denoting whether the system contains a node which has failed, and $N_{L}, N_{H}$ denoting the state of the network by represented by the lower and upper bounds of the network speeds in Mbps rounded to the closest integer. The action space is a choice between the two consensus protocols and the reward function is simply the average transaction latency of the optimised $TS$ as measured in the physical system. 

\vspace{-10pt}

\subsubsection{Results.}

\begin{figure}[h!]
\centering
\begin{subfigure}{.5\textwidth}
  \centering
  \includegraphics[width=1\linewidth]{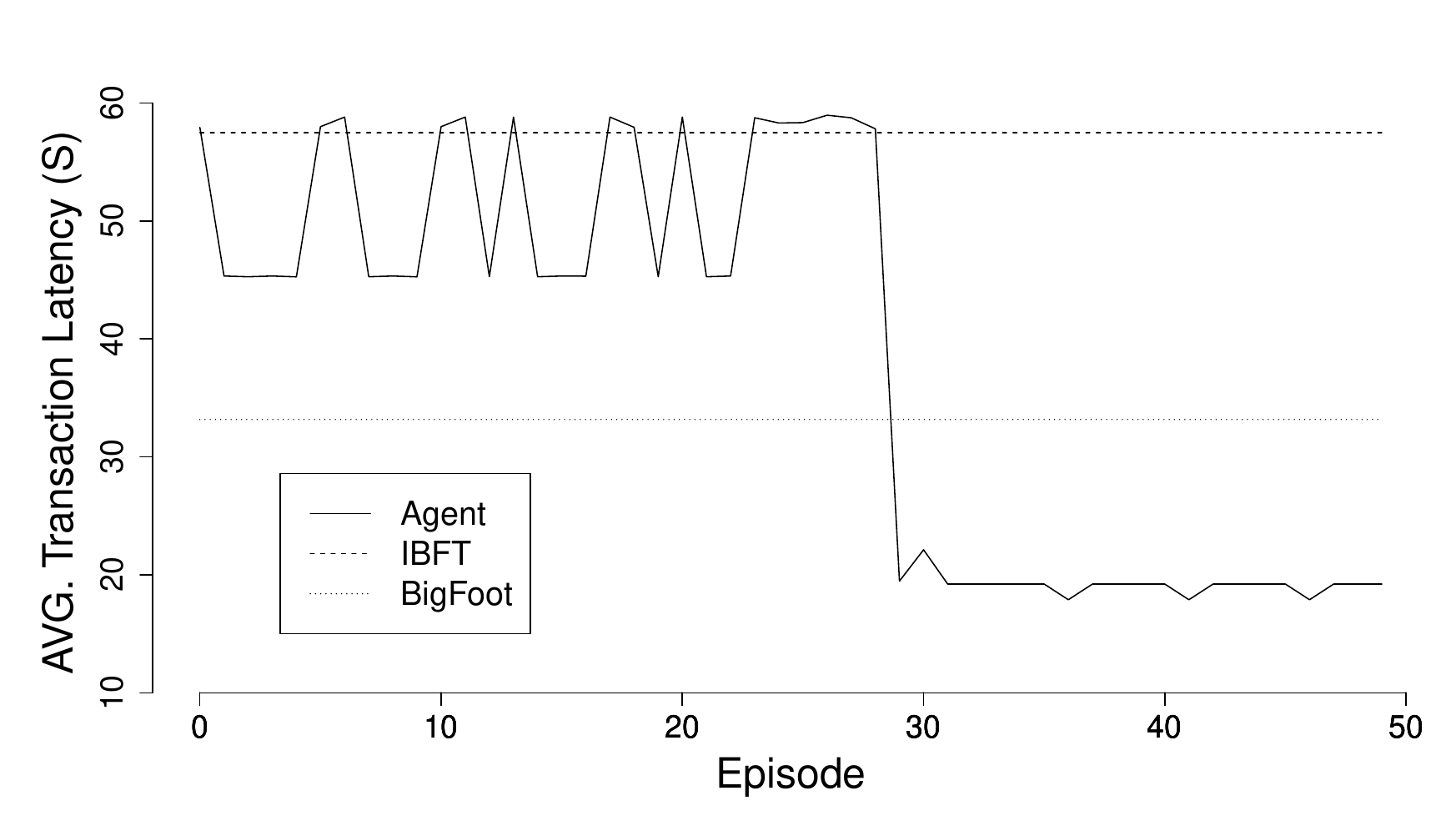}
  \caption{}
  \label{fig:sub1}
\end{subfigure}%
\begin{subfigure}{.5\textwidth}
  \centering
  \includegraphics[width=1\linewidth]{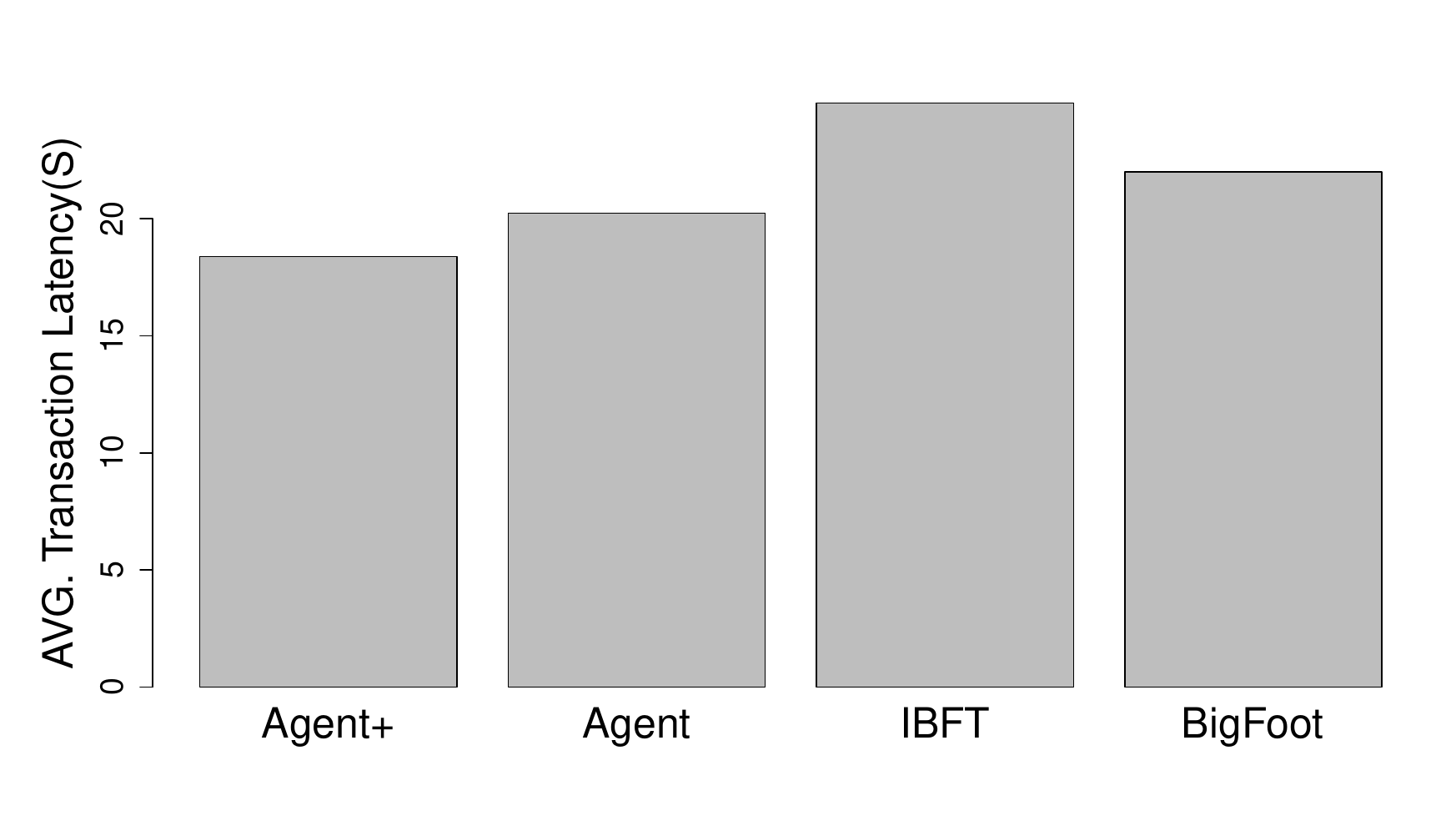}
  \caption{}
  \label{fig:sub2}
\end{subfigure}%

\caption{Results of the experimental evaluation with (a) showing the training performance of the agent on WL1 (b) the performance of the agent and the agent + simulation (denoted as agent+) for WL2} 
\label{fig:res}
\vspace{-15pt}
\end{figure}

For evaluating the performance of the proposed optimiser, the average transaction latency was used. Specifically, two workloads (WL1, and WL2) were generated using the scenario generator. WL1 was used for the training of the agent (Fig. \ref{fig:sub1}), while WL2 was used to represent the system at a later stage, where the state has evolved over time. Two approaches were used for the optimisation of WL2: (a) the agent on its own with no help from the simulator and (b) the agent augmented with simulation in the form of what-if analysis. 

\begin{figure}{}
  \centering
  \includegraphics[width=0.5\linewidth]{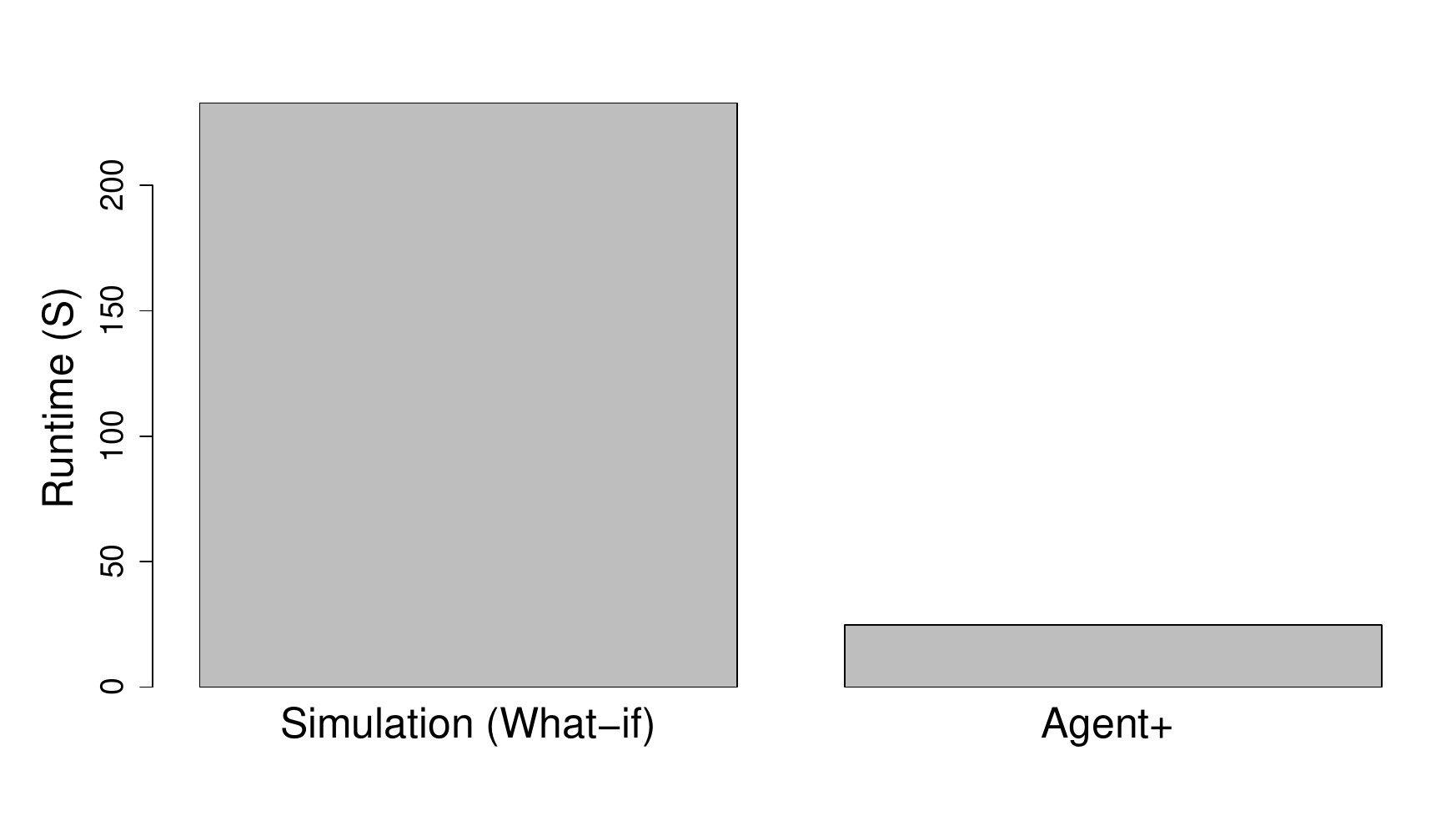}
  \caption{Comparison of the runtimes of simulation-based optimisation and agent + simulation}
  \label{fig:sub3}
  \vspace{-18pt}
\end{figure}

As shown in Fig. \ref{fig:res} the agent achieves good training performance on WL1 managing to outperform both algorithms on their own. In WL2 the agent's performance is shown to decrease in comparison to that of the agent augmented with the simulation (agent+) (Fig. \ref{fig:sub2}). Additionally, Fig. \ref{fig:sub3} shows the runtime of the agent+ as compared to that of the what-if-based optimiser demonstrating the agent's efficiency. The increased performance in combination with the reduced computational overhead of the agent+, greatly increases the potential of the proposed framework to be used in low-powered / energy-aware systems.

\section{Conclusions}
\label{Conclusions}
Leveraging on our previous work on utilising Digital Twins for dynamically managing the trilemma trade-off in blockchain systems, in this paper we have focused on the DDDAS feedback loop that links the Digital twin with the blockchain system. We have elaborated on the components and challenges to implement the loop. A key component of the feedback loop is an optimiser and we have proposed a novel optimisation approach for the system. The optimiser combines Re-enforcement Learning  and Simulation to take advantage of the efficiency of the agent with the accuracy of the simulation. Our experimental results confirm that the proposed approach not only can successfully increase the performance of the agent but do so more efficiently, requiring less computational overhead.

\section*{Acknowledgements}
This research was supported by: Shenzhen Science and Technology Program,  China (No. GJHZ20210705141807022); SUSTech-University of Birmingham Collaborative PhD Programme; Guangdong Province Innovative and Entrepreneurial Team Programme, China (No. 2017ZT07X386); SUSTech Research Institute for Trustworthy Autonomous Systems, China. 

\bibliographystyle{splncs04.bst}
\bibliography{paper.bib}

\end{document}